\documentclass[longauth]{aa} 
\usepackage{graphicx}
\usepackage{txfonts}
\begin{document}


%
\title{Discovery of VHE $\gamma$-ray emission from the BL Lac object \\
B3 2247+381 with the MAGIC telescopes}

\author{
 J.~Aleksi\'c\inst{1} \and
 E.~A.~Alvarez\inst{2} \and
 L.~A.~Antonelli\inst{3} \and
 P.~Antoranz\inst{4} \and
 M.~Asensio\inst{2} \and
 M.~Backes\inst{5} \and
 J.~A.~Barrio\inst{2} \and
 D.~Bastieri\inst{6} \and
 J.~Becerra Gonz\'alez\inst{7,}\inst{8} \and
 W.~Bednarek\inst{9}
% ...
}
\institute { IFAE, Edifici Cn., Campus UAB, E-08193 Bellaterra, Spain
 \and Universidad Complutense, E-28040 Madrid, Spain
 \and INAF National Institute for Astrophysics, I-00136 Rome, Italy
 \and Universit\`a  di Siena, and INFN Pisa, I-53100 Siena, Italy
 \and Technische Universit\"at Dortmund, D-44221 Dortmund, Germany
 \and Universit\`a di Padova and INFN, I-35131 Padova, Italy
 \and Inst. de Astrof\'{\i}sica de Canarias, E-38200 La Laguna, Tenerife, Spain
 \and Depto. de Astrof\'{\i}sica, Universidad de La Laguna, E-38206 La Laguna, Spain
 \and University of \L\'od\'z, PL-90236 Lodz, Poland
% ...
}

\abstract{}{We study the non-thermal jet emission of the BL Lac object 
B3 2247+381 during a high optical state.}
{The MAGIC telescopes  
observed the source during 13 nights between September 30th and October 30th 
2010, collecting a total of 14.2 hours of good quality 
very high energy (VHE) $\gamma$-ray 
data. Simultaneous multiwavelength data was obtained with X-ray observations
by the {{\it Swift}} satellite and optical R-band observations at the KVA-telescope. We also use high energy $\gamma$-ray (HE, 0.1\,GeV--100\,GeV) data from 
the {{\it Fermi}} satellite.}
{The BL Lac object B3 2247+381 (z=0.119) was detected, 
for the first time, at VHE 
$\gamma$-rays at a statistical significance  of 
5.6\,$\sigma$. A soft VHE spectrum 
with a photon index of -3.2 $\pm$ 0.6 was determined. 
No significant short term flux variations were found. 
We model the spectral 
energy distribution using a one-zone SSC-model, which can successfully describe our data. 
}{}

               
\titlerunning{B3~2247+381}

\maketitle
%

\section{Introduction}


The number of known extragalactic very high energy (VHE, E $>$ 100\,GeV) $\gamma$-ray emitters has increased from 6 to almost 50 (July 2011)
during the past five years\footnote{http://www.mpp.de/~rwagner/sources/}.  
Most of these objects are active galactic nuclei (AGN), especially 
blazars (flat spectrum radio quasars and BL Lac objects), which are known for their large variability
across the electromagnetic spectrum from radio to VHE $\gamma$-rays.
Eight of these new discoveries were made during high optical
 states of these sources, reported by the Tuorla blazar monitoring
program, which have triggered MAGIC observations: Mrk~180 (Albert et
al. 2006), 1ES~1011+496 (Albert et al. 2007a), BL~Lacertae (part of the data
taken as ToO due to optical high state, Albert et al. 2007b), 3C~279
(Albert et al. 2008a, Aleksi\'c et al. 2011a), MAGIC J0223+430 (Aliu et
al. 2009), S5 0716+714 (Anderhub et al. 2009),
B3~2247+381 (Mariotti et al. 2010, this paper) and
most recently 1ES~1215+303 (Mariotti et al. 2011). This suggests that
high optical states are an indication of a higher state also in 
 VHE $\gamma$-ray band, 
 at least in some sources. Additionally for PG~1553+113
a long term study (Aleksi\'c, J., Alvarez, E.A., Antonelli, L.A., Antoranz, P. et al., ApJ, in prep.) suggests a  correlation
between the optical and VHE $\gamma$-ray flux, while for PKS~2155-304 it
seems that, at least, 
in some cases the two wavebands are correlated (Foschini et al. 
2007; Aharonian et al. 2009).

The object B3 2247+381 was selected from the sample presented in 
Nieppola et al. (2006) with a reported X-ray flux F($>$ 1keV) $>2$\,mJy. 
Donato et al. (2001) classify it
as high-energy-peaked BL Lac object and the reported X-ray flux of the
source is F($>$1\,keV)= 0.6\,mJy\footnote{Note, that 
Veron-Cetty \& Veron (2006) list the source 
as a BL Lac candidate and that the classification has not been confirmed.}. 
B3~2247+381 (z = 0.119, Falco et al. 1988) has been
previously observed by the MAGIC telescope as part of the systematic
search of VHE $\gamma$-rays from X-ray bright BL Lac objects
(Aleksi\'c et al. 2011b)\footnote {The stacked dataset of all X-ray 
selected blazars resulted in 
a significant $\gamma$-ray excess, which hints towards the fact that the 
sources were emitting at a very low level.}.
MAGIC observed the source between 
August and September 2006, which resulted in an upper limit
F($>$140\,GeV) $<$ 1.6 $\cdot$ 10$^{-11}\mathrm{cm}^{-2}\mathrm {s}^{-1}$.
 The source has been monitored in the R-band by the Tuorla blazar monitoring
program ever since.

B3~2247+381 was also included in the list of potentially interesting TeV
sources released to the Imaging Atmospheric Cherenkov Telescope
experiments by the {\it Fermi}-LAT collaboration on 27th October 2009 
({\it Fermi}-LAT Collaboration 2009, priv. comm.).  
B3 2247+381 is listed in the first {\it Fermi}-LAT catalog of AGNs (Abdo et al.
2010) as 1FGL J2250.1+3825 with a very hard spectrum (spectral
index of -1.6 $\pm$ 0.1). In the second {\it Fermi}-LAT catalog this object 
is listed as 2FGL J2250+3825 with a spectral index of -1.84 $\pm$ 0.11 
(Ackerman, M., Ajello, M., Allafort, A., Antolini, E. et al., ApJ, in prep.). 
Neronov et al. (2010) found a hint of signal at 2.73\,$\sigma$ 
in the {\it Fermi}-LAT data above 100\,GeV over the period August 
2008-April 2010.



In this paper we present the first detection of VHE $\gamma$-ray
emission triggered by the optical high state of the source and the
first optical light curve of B3 2247+381. We also present simultaneous
X-ray data obtained by the {\it Swift} and High energy 
(HE) data obtained by {\it Fermi}-LAT satellites.


\section{Observations and data analysis}

%

\subsection{MAGIC observations and data analysis}  

The VHE $\gamma$-ray observations were carried out with the MAGIC 
telescopes located on the Canary Island of La Palma (28.8$^\circ$ N,
17.8$^\circ$ W at 2200 m.a.s.l). The two 17\,m telescopes use the 
atmospheric Cherenkov imaging technique and allow for 
measurements at a threshold as low as 50\,GeV.


B3 2247+381 was observed with the MAGIC telescopes during 13 nights between September 30th 
and October 30th 2010 collecting a total of 21.2 hours of data, of which 5.3 hours were 
discarded based on the event rate. 
The effective time of this observation, correcting for
the dead time of the trigger and readout systems is 14.2 hours.
Part of the data 
was taken under moderate moonlight conditions.

All the data were taken in the false-source tracking (wobble) mode (Fomin et al. 1994), in which the 
telescope was alternated every 20 minutes between two sky positions at 0.4$^\circ$ offset 
from the source. The zenith angle was between 8$^\circ$ and 35$^\circ$.  

The data was analysed using the standard MAGIC analysis framework "MARS" as 
described in Moralejo et al. (2009) with additional adaptations incorporating the stereoscopic
 observations.  
The images were cleaned using timing information as described in Aliu et al. (2009) 
with absolute cleaning levels of 6\,phe (so-called "core pixels") and 3\,phe 
("boundary pixels") for the first telescope and 9\,phe and 4.5\,phe respectively for
 the second telescope. The images were parameterized in each telescope separately following the prescription of Hillas (1985).

In order to reconstruct the shower arrival direction we used the
random forest regression method (RF DISP method, Aleksic et al.
2010) which was extended using stereoscopic information such as the
height of the shower maximum and the impact distance of the shower on
the ground. We estimated the RF DISP for each telescope separately and
obtained two possible solutions along the major axis of the shower
image in each telescope, respectively. Finally, we searched for the
combination of two solutions (one from each telescope) that have the
shortest squared angular distance between them. If this squared
angular distance is greater than 0.05 degree$^{2}$ the event is
removed from further analysis. This improves the background rejection
since hadron induced showers have a larger error on the reconstruction
of the arrival direction. The final arrival direction is the mean of
the two solutions (weighted by the number of pixels in each shower
image).



For the gamma-hadron separation the random forest method is used (Albert et al. 2008b). 
In the stereoscopic analysis image parameters of both telescopes are used as well 
as the shower impact point and the shower height maximum. 
A detailed description of the stereoscopic MAGIC analysis can be found in
(Aleksic J., Alvarez, E.A., Antonelli, L.A., Antoranz, P. et al., Astr. Particle Phys. in prep.).

\subsection{Optical observations and data analysis}

B3 2247+381 has been observed by the Tuorla group\footnote{http://users.utu.fi/kani/},  using the  
35\,cm KVA telescope, located on La Palma, since 2006 (see Aleksi\'c et al. 2011a for a description of the telescope). Observations 
have been made in the R-band. The brightness of the object 
was measured based on stars in the same CCD-frames as B3 2247+381. 
These stars were calibrated by the Tuorla group during the observing seasons.
R-band magnitudes were converted to fluxes, 
using: F(Jy)= $3080 \cdot 10^{-\mathrm{m_{R}}/2.5}$.
The fluxes were corrected for galactic absorption by R=0.398\,mag 
(Schlegel et al. 1998). 
During the years 2006-2009 B3 2247+381 was a quite faint and steady source 
at R $\sim$1.8\,mJy, but during late summer 2010 it went to a high optical 
state, 
reaching an average flux level of 2.4\,mJy. 
The source also stayed at this level throughout the observing 
season (see Figure 3). 
In late September an alert was given to MAGIC and it started observations 
on September 30th. 
An alert is issued when the optical flux has increased by 50\% from the 
long term running average. 

\subsection{Swift observations and data analysis}

The prime objective of the \textit{Swift} Gamma-Ray Burst observatory, launched in November 2004 (Gehrels et al. 2004),
was to detect and follow up on Gamma-Ray bursts, but it has turned into a 
multi-purpose observatory 
due to its fast slewing and response capacity and its multi-wavelength coverage. \textit{Swift} is equipped with three 
telescopes, the Burst Alert Telescope (BAT; Barthelmy 2005),  
which covers the 15-150\,keV range, the X-ray telescope 
(XRT; Burrows et al. 2005) 
covering the 0.3-10\,keV energy band, and the UV/Optical Telescope (UVOT; 
Roming et al. 2005) covering the 1800-6000\,\r{A} wavelength range. 

\textit{Swift} Target of Opportunity observations were requested and from 
October, 
5 to 16, 2010. \textit{Swift}/XRT observed the source for $\sim$ 5\,ks every night, in Photon counting (PC) mode.
 We also analysed Swift archival data from August 10th, 2009, February 18th, 2010 and  April 18th, 2010 in order 
to compare the  level of the X-ray emission to previous observations. 

The data were processed with standard procedures using the FTOOLS task XRTPIPELINE (version 0.12.6) distributed by 
HEASARC within the HEASoft package (v.6.10). Events with grades 0-–12 were 
selected for the PC data (see Burrows
 et al. 2005) and the response matrices available in the \textit{Swift} CALDB (v.20100802) were used. For the 
spectral analysis, we extracted the PC source events in the 0.3-10\,keV range within a circle with a radius of 20 
pixels ($\sim 47\,$ arcsec). The background was extracted from an off-source circular region of 40-pixels radius.

The spectra were extracted from the corresponding event files and binned using GRPPHA to ensure a minimum of 30 counts 
per bin in a manner so that the $\chi ^2$ statistic could reliably be used.  
Spectral analyses were performed using XSPEC version 12.6.0.
We adopted both a simple power law model and a log-parabolic model as in 
Massaro et al. (2004) 
with an absorption hydrogen-equivalent column density fixed to the Galactic value in the 
direction of the source, namely $1.2\times 10^{21}\, \rm{cm^{-2}}$. The two models provide similar results in terms of 
goodness of fit above $\sim 0.7\,$keV. However below this energy the 
differences are in general negligible due to low statistics.\\


\textit{Swift}/UVOT observed the source in the "filter of the day" mode, that is a different filter was used for 
different observations. This does not allow to compare the UV fluxes among different days. UVOT  source counts were extracted 
from a circular region 5 arcsec-sized centered on the source position, while the background was extracted from a larger 
circular nearby source–free region. 
These data were processed with the {\tt uvotmaghist} task
 of the HEASOFT package.
The observed magnitudes have been corrected for Galactic extinction $E_{B-V}=0.149$\,mag (Schlegel et al. 1998).


\subsection{Fermi data analysis}  

The {\it Fermi}-LAT is a pair conversion telescope designed to cover
the energy band from 20\,MeV to greater than 300\,GeV (Atwood
et al. 2009) which operates in survey mode, scanning the entire
sky every three hours.
The data used in this paper encompasses the time interval
from August 5th, 2008 to April 7th, 2011 (MJD 54683 -
55658), and were analyzed with the {\it Fermi} Science Tools package version
v9r23p0, which are available from the {\it Fermi} Science
Support Center (FSSC).
Only events with the highest probability of being photons,
those in the diffuse class, located within 12$^\circ$ of B3 2247+304 were
used in this analysis. In addition, a cut on the maximum zenith
angle ($<$ 100$^\circ$) was applied to reduce the contamination from
the Earth-limb gamma-rays, which is produced by cosmic-rays interacting
with the upper atmosphere.
The background model used to extract the $\gamma$-ray signal includes a 
Galactic diffuse emission component and an isotropic component 
(including residual instrument background), modelled with the files
 gll\_iem\_v02\_P6\_V11\_DIFFUSE.fit
and isotropic iem\_v02\_P6\_V11\_DIFFUSE.txt, which are publicly
available\footnote{http://fermi.gsfc.nasa.gov/ssc/data/access/lat/BackgroundModels.html}. 
The normalizations of the components comprising
the total background model were allowed to vary freely during
the spectral point fitting.
The spectral fluxes were derived with the post-launch instrument
response functions (IRF) P6\_V11\_DIFFUSE, 
and applying an unbinned maximum likelihood technique (Mattox et al. 1996) to
events in the energy
range spanning from 300\,MeV to 300\,GeV. All the sources from
the 2FGL catalog (Abdo, A.A., Ackermann, M., Ajello, M., Allafort, A. et al., ApJS, in prep.) located within 7$^\circ$ radius of B3
2247+38 were included in the model of the region. The initial
parameters in the XML file were set to those of the 2FGL
catalog, leaving the normalization parameters free in the fitting
procedure. The systematic uncertainty in the flux is estimated
as 10\% at 100\,MeV, 5\% at 560\,MeV and 20\% at 10\,GeV and
above\footnote{http://fermi.gsfc.nasa.gov/ssc/data/analysis/LAT\_caveats.html}.

For the period of the MAGIC observations (30 days between
September 30th and October 30th 2010), the source is not significantly
resolved, and hence only 95\% confidence level upper limits were obtained.
In the light curve presented in Fig.\ 3, for each time bin, if the TS value 
for the source was TS $<$ 4 or the number of predicted photons $N\_pred <$3, 
the values of the fluxes were replaced 
by the 95\% confidence level upper limits.
The 2-sigma upper limits were computed using the Bayesian method
(Helene 1983), where the likelihood is integrated from 0 up to the
flux that encompasses 95\% of the posterior probability.

   \begin{figure}
   \includegraphics[width=9.0cm]{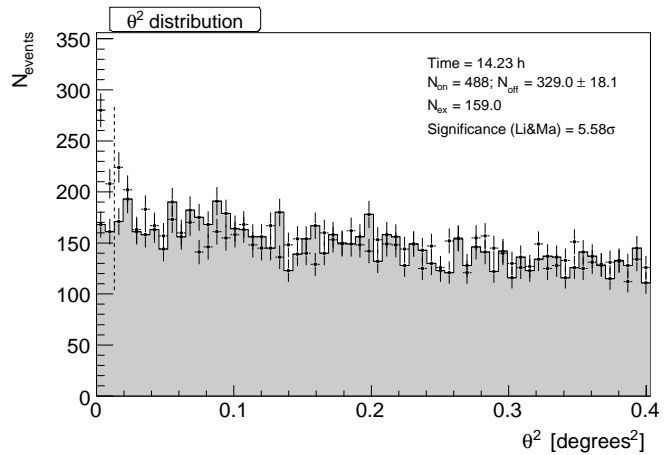}
 \caption{Distribution of the squared angular distance ($\theta^2$) for  
the on-source counts in the direction of B3 2247+381 (black 
points with errorbars) and the normalized off-source events (gray histogram).}
    \end{figure}

   \begin{figure}
   \includegraphics[width=9.0cm]{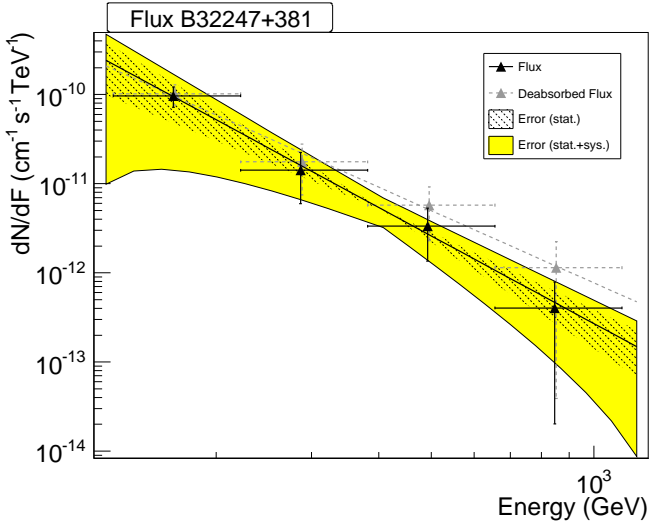}
\caption{The unfolded differential energy spectrum of B3 2247+381 
observed by MAGIC. The black data points refer to the measured spectrum, 
while the grey dashed points have been corrected for the attenuation of 
the EBL. The thick black and dashed grey lines are power-law fits to the respective data points 
(fit results are given in the text). The dashed band corresponds to the 
statistical error of the fit to the measured spectrum, while the white band surrounding it is the sum of the statistical and systematic errors of the fit.}

    \end{figure}

   \begin{figure}
   \includegraphics[width=14.0cm, angle=270]{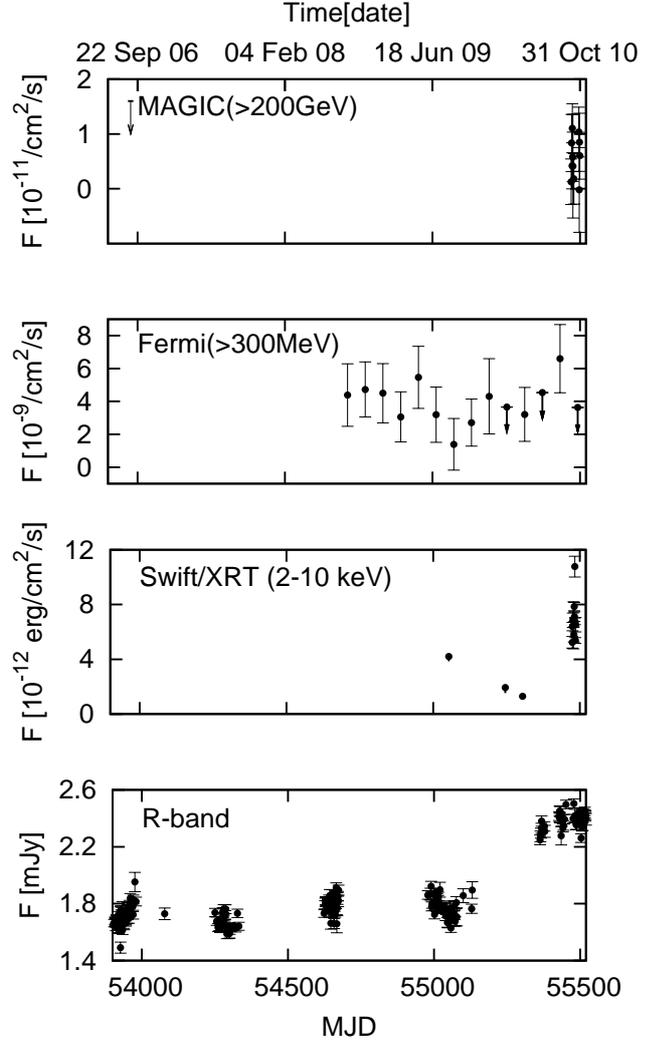}
 \caption{Long term light curves of B3 2247+381 in VHE $\gamma$-rays, 
{\it Fermi}-LAT  
HE $\gamma$-rays (two months time intervals), {\it Swift}
X-rays and optical KVA R-band.}
    \end{figure}

   \begin{figure}
   \includegraphics[width=14.0cm, angle=270]{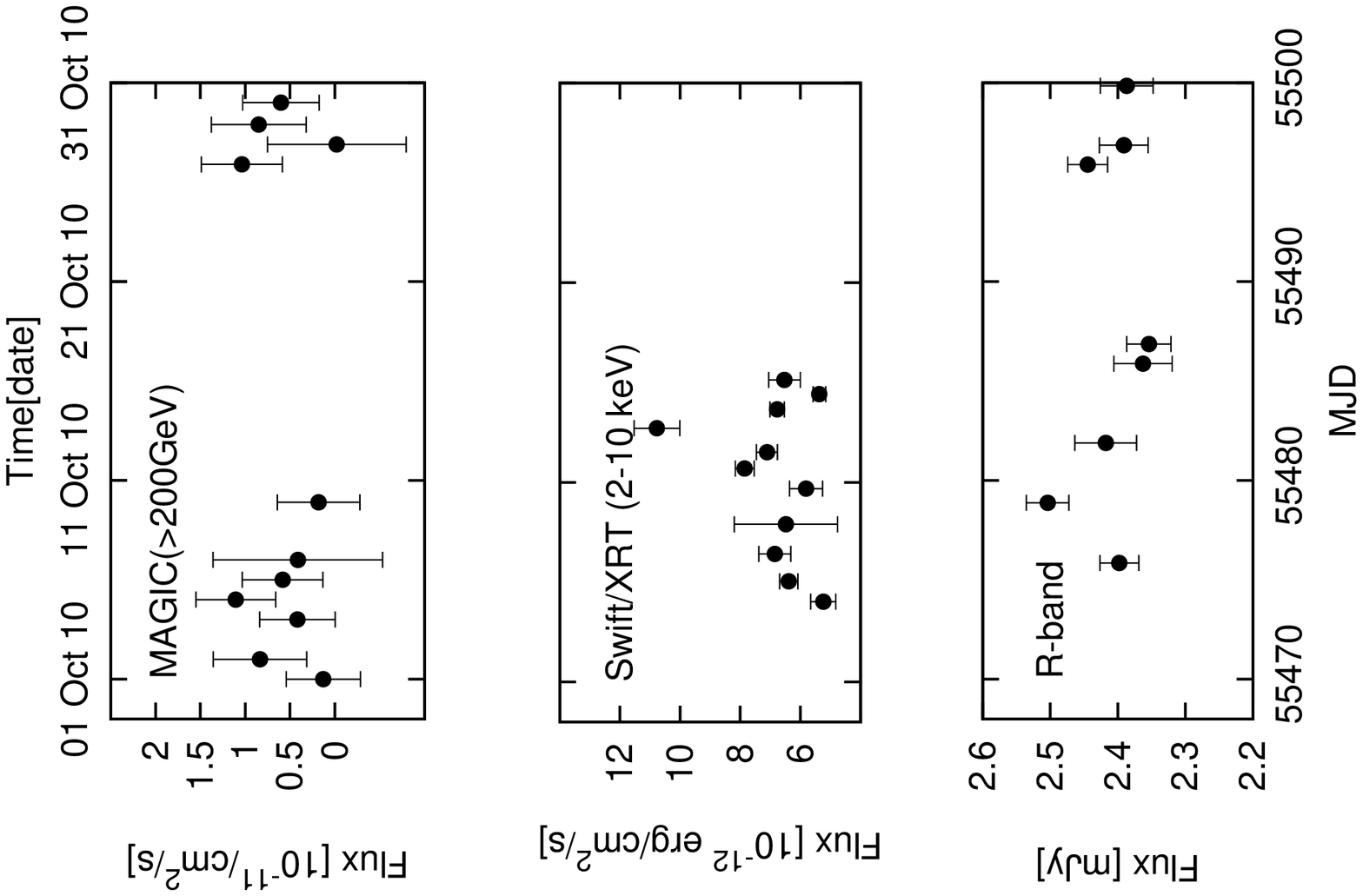}
 \caption{Same as Fig. 3, but zoomed into the time interval of the MAGIC 
observations in September-October 2010. }
    \end{figure}

\section{Results}

In the MAGIC data the distribution of the square of the angle $\theta$ between the reconstructed direction of the events after cuts and the position of B3 2247+381 (RA: 22.83472$^\circ$, DEC: 38.41028$^\circ$, J2000 as in Ficarra et al.\ 1985) shows an excess of 159 $\pm$ 28 events above a threshold of 200\,GeV. 
The cuts were previously optimized on a trial sample of Crab Nebula data. 
The threshold  was calculated from Monte Carlo simulated data by
finding the peak of the differential rate vs. 
energy distribution after cuts and spectral re-weighting. 
The expected background level with the same cuts is 329 events, 
calculated from the $\theta^2$ distribution of the reconstructed direction of the events with respect to the anti-source position, located 180$^\circ$ from the real source position in the camera plane (Figure 1). 
The measured excess corresponds to a post-trial significance of 5.6\,$\sigma$ calculated using eq. 17 from Li and Ma (1983).

The source position and extension, determined by a 2D Gaussian 
fit of the sky map produced with the cuts above, are 
consistent with a point-like source placed at the position
 of B3 2247+381 within 0.015$^\circ$, well 
within the statistical uncertainty and the systematic pointing
 uncertainties of MAGIC. 


The integral flux of the source above 200\,GeV was estimated to be 
$(5.0 \pm 0.6_{\mathrm{stat}} \pm 1.1_{\mathrm{sys}})\cdot 10^{-12}$ ph cm$^{-2}$s$^{-1}$. 
The differential energy spectrum is well described by a simple power-law: 
$dN/dE = f_0$ (E/300\,GeV)$^{\gamma}$. 
The photon index $\gamma$ was found to be $-3.2 \pm 0.5_{\mathrm{stat}} \pm 0.5_{\mathrm{sys}}$, 
and a flux normalization $f_0$ at 300\,GeV of  
$(1.4 \pm 0.3_{\mathrm{stat}} \pm 0.2_{\mathrm{sys}})\cdot 10^{-11}$ ph cm$^{-2}$s$^{-1}$TeV$^{-1}$. 
In order to correct the effects in the spectrum determination introduced by the 
limited energy resolution, different unfolding algorithms 
(Forward, Tikhonov, Schmelling 1\&2, Bertero; all described in 
Albert et al. (2007c)) were used, and all agreed within errors. 
Taking into account the attenuation due to pair production with the extragalactic 
background light (EBL), the spectrum was found to be compatible with a power law with photon 
index $\gamma = -2.7 \pm 0.5_{\mathrm{stat}} \pm 0.5_{\mathrm{sys}}$ 
and flux at 300\,GeV $f_0$ = $(2.0 \pm 0.3_{\mathrm{stat}} \pm 0.3_{\mathrm{sys}})\cdot 10^{-11}$ 
ph cm$^{-2}$s$^{-1}$TeV$^{-1}$ (Figure 2). 
Two different EBL models were used (Dom\'inguez et 
al. 2011 and  Kneiske \& Dole 2010), 
and they were found to be in good 
agreement with each other, well within the statistical uncertainties. 

Long term light curves of B3 2247+381 in VHE $\gamma$-rays (MAGIC), HE $\gamma$-rays ({\it Fermi}-LAT),
X-rays ({\it Swift}) and optical (Tuorla Observatory) are shown in Figure 3. 
Our detection in VHE $\gamma$-rays is compatible with the previous 
upper limit from 2006 and thus no 
variability can be established in this energy band. 
However, in X-rays and the optical band a clear increase of the flux in fall 2010 is 
evident, while the {\em Fermi}-LAT light curve is consistent with a constant flux.  A fit with a constant to the eleven {\em Fermi} flux points where the source is significantly detected, gave a flux value of $(3.7 \pm 0.5)\cdot 10^{-9}$ ph cm$^{-2}$s$^{-1}$, with a reduced $\chi^2$ of 0.7 with ten degrees of freedom. The {\em Fermi}-LAT is not sensitive enough for detecting short term variations at this 
flux level.
The temporal evolution of the VHE $\gamma$-ray, 
X-ray and optical 
emission from B3~2247+381 
during  September-October 2010 
observation shows no strong variability on time scales of a 
night (Figure 4). 
In particular, the MAGIC light curve above 200\,GeV is consistent with a non-variable
source, having a reduced $\chi^{2}$ of $0.6$ with ten degrees of freedom. During one night the X-ray flux is significantly higher 
(almost factor 2) than the other X-ray points, but unfortunately 
we do not have 
simultaneous optical or MAGIC data for that night.  

%

\section{Discussion}

In this paper we report the  discovery of VHE $\gamma$-rays from
B3~2247+381 by MAGIC. The MAGIC observations were triggered
by an optical high state of the source, like several other
discoveries. However, the observed VHE $\gamma$-ray flux is consistent
with the upper limit from previous observations and we therefore cannot
conclude if the source was in a higher VHE $\gamma$-ray state during the
observations. 


In Figure 5 we show the spectral energy distribution of the source during the 
MAGIC observations, together with simultaneous {\it Swift} and optical data, and other
 non-simultaneous data. The {\it Swift} observations show that the source was in a high state also in X-rays.  
In the {\it Fermi} energy range the source is very weak, which limits the capability of detecting statistically significant flux-variability on time scales of a few months. 
The synchrotron component of the SED is showing a significantly larger emission in the high state, while the inverse Compton component is consistent with only minor changes. We must however note that the weak detection in the HE and VHE $\gamma$-ray band significantly limits the determination of the inverse Compton component. 
 
We reproduce the SED with a one-zone synchrotron-self-Compton (SSC) model (see  
Tavecchio et al. 2001 for a description). 
In brief, the emission region
is assumed to be spherical, with a radius $R$, filled with a tangled
magnetic field of intensity $B$. The relativistic electrons follow a
smoothed broken power-law energy distribution specified by the limits
$\gamma _{\rm min}$, $\gamma _{\rm max}$ and the break at 
$\gamma_{\rm b}$ as well as the slopes $n_1$ and $n_2$ before and after the break, respectively. 
Relativistic effects are taken into account by the Doppler
factor $\delta$. The used input model parameters are shown in Table 1.

To reproduce the change of the Compton and synchrotron luminosity ratio between the low and the high state we mainly act on the electron normalization, source radius and Doppler factor (with slight changes also to the other parameters). The steeper X-ray slope in the low state implies a larger value of $n_2$. 

The comparison with parameters derived for BL Lac objects (e.g.\ Tavecchio et al. 2010) reveals that the parameters used for B3 2247+381 are close to the typical values. As for other sources, the somewhat larger (lower) value of the Doppler factor $\delta$ (the magnetic field intensity $B$) with respect to ``standard" values is mainly due to the relatively large separation between the synchrotron and IC peaks.

Additionally, we model the SED, during the high state, 
with the one-zone SSC code from Weidinger et al. (2010) 
which is shown in Figure 6. 
In contrast to the model from Tavecchio et al. (2001) all parameters are basic
physical parameters and the electron and photon spectrum and their breaks
are derived self-consistently with a continuous injection of monochromatic electrons at the energy $\gamma_0=10^4$ and injection rate $K=8.4\cdot 10^4$ cm$^{-3}$ s$^{-1}$. The spectrum is the resulting steady-state solution. The environment is defined by the magnetic field $B=0.07$\,G, the acceleration
efficiency $t_{acc}/t_{esc}=1.09$ (i.e.\ the particle spectral index is $s=2.09$
and the resulting $\gamma_{max}=4.8\cdot 10^5$), and the blob radius
$R=1.3\cdot 10^{16}$\,cm. The break in the e$^-$ spectrum of 1 at
$\gamma_b=2.9\cdot 10^4$ arises self-consistently from IC and synchrotron
cooling. The common parameters of both models agree
very well.


   \begin{figure}
   \includegraphics[width=9.0cm]{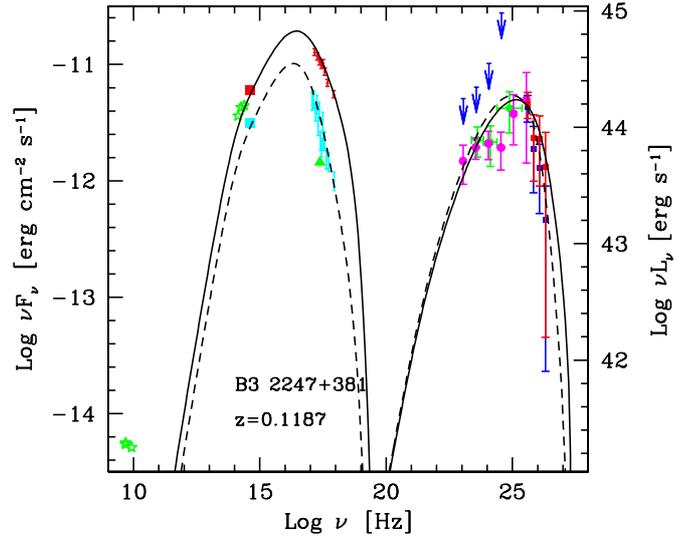}
 \caption{Spectral energy distribution of B3 2247+381 (red: EBL corrected MAGIC spectral points). 
The green crosses are 
1FGL {\it Fermi} data points (Abdo et al. 2010), while the pink points represent the {\it Fermi} analysis from this work (2.5 years of data). Blue arrows show the 95\% confidence upper limits computed from {\it Fermi}-LAT data for the time interval of the MAGIC observation. 
Low (high) state {\it Swift} data were taken on April 18th 2010 (October 5-16, 2010). 
The host galaxy contribution has been subtracted from the KVA R-band data (red and light blue squares), following Nilsson et al. (2007). The data have been corrected for galactic absorption. 
Green and light blue points represent non-simultaneous low state data. 
The solid line is our SSC-model fit to the high state observations; 
the dotted line is a fit to the low state observations.}
    \end{figure}


\begin{table*}
\caption{Input parameters for the high and low states of the SSC-model shown as solid and dashed lines in Fig. 5. For more explanations see text.} 
\label{table:1} 
\centering 
\begin{tabular}{c c c c c c c c c c} 
\hline\hline 
Flux State & $\gamma_{\rm min}$ & $\gamma_{\rm b}$ & $\gamma_{\rm max}$ & $n_1$ & $n_2$ & $B$ & $K$ & $\delta$ & $R$ \\ 
 &  &  &  &  &  & G & $cm^{-3}$ &  & cm \\ 
\hline 
High & 3 $\cdot 10^3$ & $7.1 \cdot 10^4$ & 6 $\cdot 10^5$ & 2.0 & 4.35 & 0.06 & 2.5 $\cdot 10^3$ & 35 & $8\cdot10^{15}$\\ 
Low  & 3 $\cdot 10^3$ & $6.8 \cdot 10^4$ & $5 \cdot 10^5$ & 2.0 & 5.35 & 0.08 & $1.15 \cdot 10^4$& 30 & $4\cdot10^{15}$\\
\hline 
\end{tabular}
\end{table*}

   \begin{figure}
   \includegraphics[width=7.0cm, angle=270]{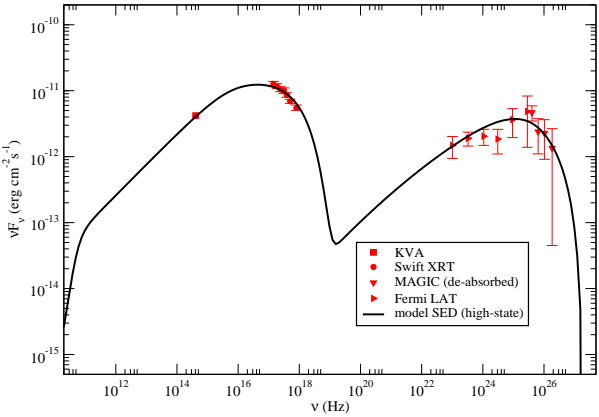}
 \caption{The solid line shows the SED model using Weidinger et al. (2010). The data points are described in the inlay of the figure. The fit parameters can be found in the text.}
    \end{figure}

The optical monitoring of candidate VHE $\gamma$-ray blazars has proven to be a successful tool for their discovery. However, for B3~2247+381 (like for Mrk~180 and 1ES~1011+496) we cannot establish a connection between the optical high state and the VHE $\gamma$-ray emission since the upper limit from previous observations (during a low optical state) is higher than the detected VHE $\gamma$-ray flux during the discovery. Therefore further observations are still needed to study the connection between these two wavebands (see also Reinthal et al. 2011).

\begin{acknowledgements}

We would like to thank the Instituto de Astrof\'{\i}sica de
Canarias for the excellent working conditions at the
Observatorio del Roque de los Muchachos in La Palma.
The support of the German BMBF and MPG, the Italian INFN, 
the Swiss National Fund SNF, and the Spanish MICINN is 
gratefully acknowledged. This work was also supported by 
the Marie Curie program, by the CPAN CSD2007-00042 and MultiDark
CSD2009-00064 projects of the Spanish Consolider-Ingenio 2010
programme, by grant DO02-353 of the Bulgarian NSF, by grant 127740 of 
the Academy of Finland, by the YIP of the Helmholtz Gemeinschaft, 
by the DFG Cluster of Excellence ``Origin and Structure of the 
Universe'', and by the Polish MNiSzW Grant N N203 390834.
The $Fermi$-LAT Collaboration acknowledges support from a number of agencies and institutes for both development and the operation of the LAT as well as scientific data analysis. These include NASA and DOE in the United States, CEA/Irfu and IN2P3/CNRS in France, ASI and INFN in Italy, MEXT, KEK, and JAXA in Japan, and the K.~A.~Wallenberg Foundation, the Swedish Research Council and the National Space Board in Sweden. Additional support from INAF in Italy and CNES in France for science analysis during the operations phase is also gratefully acknowledged.
We gratefully acknowledge N. Gehrels for approving this set of ToOs and
the entire Swift team, the duty scientists and science planners for the
dedicated support, making these observations possible.

\end{acknowledgements}

\end{document}